\documentclass[a4paper, 11pt]{article}

\usepackage{graphicx}
\usepackage{epsf,amsmath,bbold,amsfonts,stmaryrd}

\usepackage[utf8]{inputenc}
\usepackage{mathrsfs}
\usepackage{appendix}
\usepackage{amssymb}
\usepackage{float}
\usepackage{color}
\usepackage{cite}
\usepackage{hyperref}
\hypersetup{pageanchor=false}
\usepackage{indentfirst}
\usepackage{url}
\usepackage{float}
\usepackage{caption}
\usepackage[numbers,square,comma,sort&compress,merge]{natbib}

\usepackage{subfigure}

\usepackage{ulem}

\def\nn{\nonumber}

\def\be{\begin{equation}}
\def\ee{\end{equation}}

\def\bea{\begin{eqnarray}}
\def\eea{\end{eqnarray}}

\def\ba{\begin{array}}
\def\ea{\end{array}}

\def\bc{\begin{center}}
\def\ec{\end{center}}

\def\bl{\begin{flushleft}}
\def\el{\end{flushleft}}

\def\br{\begin{flushright}}
\def\er{\end{flushright}}

\def\bi{\begin{itemize}}
\def\ei{\end{itemize}}

\def\bt{\begin{tabular}}
\def\et{\end{tabular}}

\newcommand{\xmi}{x_\text{min}}
\newcommand{\xma}{x_\text{max}}
\newcommand{\ymi}{y_\text{min}}
\newcommand{\yma}{y_\text{max}}

\numberwithin{equation}{section}

\begin{document}
\title{Is a photon ring invariably a closed structure?}

\author{
Xiangyu Wang$^{1}$, Xiaobao Wang$^{2}$, Hai-Qing Zhang$^{3, 4}$, Minyong Guo$^{1, 5\ast}$}
\date{}

\maketitle

\vspace{-10mm}

\begin{center}
{\it
$^1$ Department of Physics, Beijing Normal University,
Beijing 100875, P. R. China\\\vspace{4mm}

$^2$ School of Applied Science, Beijing Information Science and Technology University, 
Beijing 100192, P. R. China\\\vspace{4mm}

$^3$ Center for Gravitational Physics, Department of Space Science,
Beihang University, Beijing 100191, China\\\vspace{4mm}

$^4$ Peng Huanwu Collaborative Center for Research and Education,
Beihang University, Beijing 100191, China\\\vspace{4mm}

$^5$ Key Laboratory of Multiscale Spin Physics, Ministry of Education, 
Beijing 100875, P. R. China\\\vspace{4mm}
}
\end{center}

\vspace{8mm}

\begin{abstract}

 In this study, we investigate the image of a rotating compact object (CO) illuminated by a geometrically thin, optically thin disk on the equatorial plane. As the radius of the CO's surface fluctuates, the CO may partially or entirely obscure the photon region. We observe that the perceived photon ring may exhibit discontinuities, deviating from a closed structure, and may even disappear entirely. We find that the disruption and disappearance of the photon ring are dependent on the observational angle—a novel phenomenon not previously observed in black hole imaging studies. Our study reveals that while the factors influencing this unique photon ring phenomenon are diverse and the outcomes complex, we can provide a clear and comprehensive explanation of the physical essence and variation trends of this phenomenon. We do this by introducing and analyzing the properties and interrelationships of three characteristic functions, $\tilde{\eta}$, $\eta_o$, and $\eta_s$  related to the photon impact parameters. Additionally, our analysis of the intensity cuts and inner shadows of the images uncovers patterns that differ significantly from the shadow curve.

\end{abstract}

\vfill{\footnotesize $\ast$ Corresponding author: minyongguo@bnu.edu.cn}

\maketitle

\newpage
\baselineskip 18pt
\section{Introduction}\label{sec1}

In recent years, the Event Horizon Telescope (EHT) has successively revealed intensity and polarized images of M87* and SgrA* \cite{EventHorizonTelescope:2019dse, EventHorizonTelescope:2022wkp, EventHorizonTelescope:2021bee}, and more recently, has made significant progress in multi-wavelength observations as well \cite{EventHorizonTelescope:2022apq, Lu:2023bbn}. A salient feature in these images is the existence of a luminous ring \cite{EventHorizonTelescope:2019pgp}. Theoretical investigations suggest that this observed feature not only provides direct evidence for general relativity in strong field regimes but also robustly substantiates the existence of black holes \cite{EventHorizonTelescope:2019ggy, EventHorizonTelescope:2022xqj, Younsi:2021dxe, Glampedakis:2021oie}. Nonetheless, some studies propose that due to the current constraints in the EHT's resolution, the photon ring cannot be discerned from the existing images \cite{EventHorizonTelescope:2019jan}. Furthermore, compact objects (COs) without horizons could potentially produce images similar to those of black holes \cite{Rosa:2022tfv, Rosa:2022toh, Herdeiro:2021lwl, Sengo:2024pwk, Vincent:2020dij}, characterized by a bright annular region and a central dark area, commonly referred to as the inner shadow \cite{Chael:2021rjo, Hou:2022eev, Zhang:2024lsf, Chen:2024nua}. 

In the investigation of black hole shadows and images, three closely related concepts have often been intertwined in previous research: the shadow curve, the critical curve, and the photon ring \cite{Gralla:2019xty, Hu:2020usx, Hou:2021okc}. In the vicinity of a black hole or a horizonless compact object, strong gravitational fields create a shadow region in the observer's field of view when a light source is present. This shadow's boundary is known as the shadow curve. However, the appearance of the shadow can vary for the same compact object, depending on the nature of the light source. The first type of light source involves background light from galaxies or a spherical light source surrounding the compact object \cite{Perlick:2021aok, Hu:2020usx}. In this case, we typically refer to the boundary as the shadow curve. The second type of light source involves accretion flows or outflows, where the inner region does not emit light due to the existence of an inner boundary. This inner region corresponds to a shadow, and the image of the inner boundary is also referred to as the shadow curve. However, it is more apt to call this the inner shadow \cite{Chael:2021rjo, Hou:2022eev}. Thus, unless otherwise specified, the term shadow curve in this context specifically refers to the scenario corresponding to the first type of light source. 

\begin{figure}[h!]
\centering
\includegraphics[width=0.4\textwidth]{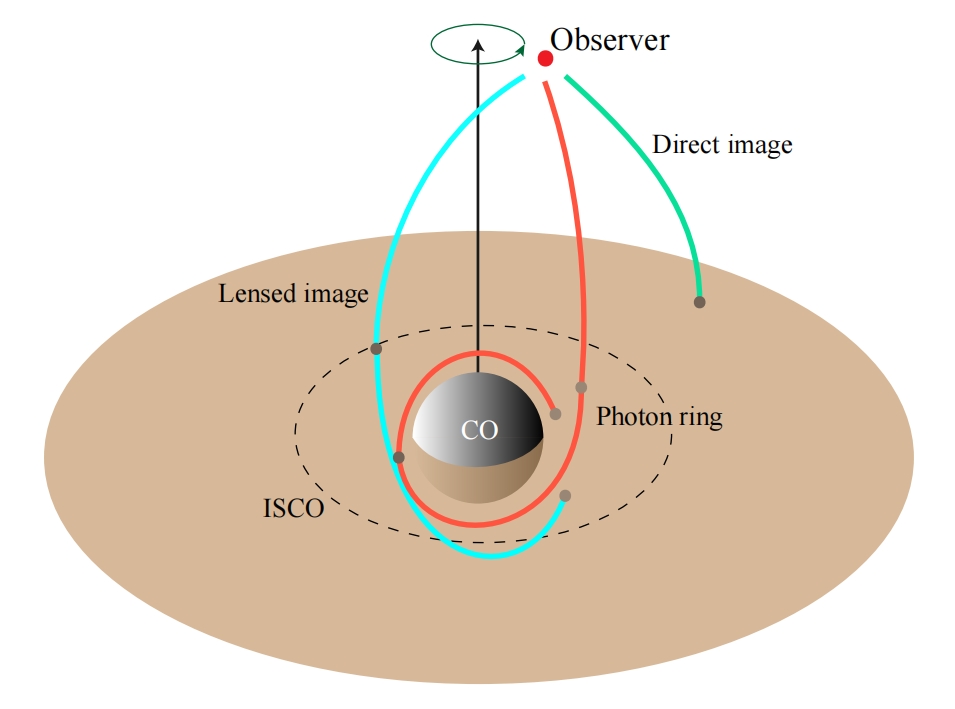}

\caption{The green curve depicts a orbit of photons producing direct images, while the red curve represents lensing images, and the blue curve indicates photon images. A dashed curve, marked as ISCO, corresponds to the innermost stable circular orbits.}
\label{cst}
\end{figure}

In the context of the critical curve, as defined in \cite{Gralla:2019xty}, this term refers to the image of photons that, when traced backward from the perspective of the observer, asymptotically approach bound photon orbits. The aggregate of all such bound photon orbits constitutes what is known as the photon region, or the photon sphere in the case of spherically symmetric spacetimes. It's worth noting that when the central celestial body is a black hole, the shadow curve and the critical curve coincide. However, for a compact object without an event horizon, the shadow curve and the critical curve do not fully overlap \cite{Wang:2023nwd}. 

The definition of a photon ring has varied in previous literature. In references \cite{Johnson:2019ljv, Gralla:2019drh, Hou:2022gge}, all images beyond the primary image are collectively referred to as the photon ring. Conversely, in references \cite{Gralla:2019xty, Zhang:2024lsf, Huang:2024wpj}, the photon ring label is applied to images beyond both the primary and secondary images. We adopt the latter definition for this study. Notably, the critical curve, by definition, is a component of the photon ring, as it can be conceptualized as an infinite-order image. Therefore, in many earlier works, the distinction between the photon ring and the critical curve was not highlighted. Additionally, in black hole research, the critical curve aligns with the shadow curve, often collectively referred to as the black hole shadow \cite{Cunha:2018acu, Grenzebach:2014fha, Vagnozzi:2019apd, Amarilla:2011fx, Wei:2013kza, Guo:2020zmf, Perlick:2018iye, Li:2013jra, Zeng:2020dco, Li:2020drn, Zeng:2020vsj, Wang:2017hjl, Hou:2018avu, Chen:2022nbb, Gan:2021pwu, Wang:2018eui, Kuang:2022xjp}.

In our previous work \cite{Wang:2023nwd}, we found that for a rotating CO, the shadow curve and the critical curve do not completely overlap; a portion of the critical curve is encompassed within the shadow curve, suggesting that some parts of the critical curve are unobservable. We propose that if the light source comes from an accretion disk, the observed photon ring might show discontinuities or could even disappear, a scenario less likely with a black hole. Specifically, we consider a geometrically thin and optically thin accretion disk model, as shown in Fig. \ref{cst}. We determine the order of the images by the number of intersections between the light rays reaching the observer and the accretion disk: a single intersection represents the primary image, two intersections denote the secondary image, and three or more intersections indicate the photon ring. These are color-coded as light green, light blue, and red lines, respectively, in Fig. \ref{cst}. This model has been extensively used in previous studies, as exemplified by references \cite{Chael:2021rjo, Hu:2023pyd, Zeng:2023fqy, Meng:2023htc, Cao:2024kht, Hamil:2024ppj, DeMartino:2023ovj, Huang:2023ilm, Zhang:2023okw, Huang:2023yqd, Guerrero:2022msp, Rosa:2022tfv, Chakhchi:2022fls, Guerrero:2021ues, Li:2021riw} for spherically symmetric spacetimes and references \cite{Gralla:2019xty, Hou:2022eev, Wang:2023fge} for axially symmetric spacetimes. We anticipate observing instances of disrupted or vanished photon rings. Our investigation will consider not only the position of the CO's surface radius but also the observational angle, indicating a complex correlation. Subsequently, we will introduce three characteristic functions related to the photon impact parameters: $\tilde{\eta}$, $\eta_o$, and $\eta_s$ to provide a comprehensive explanation. Furthermore, for the sake of comprehensiveness, we also plan to investigate the intensity cuts and inner shadows of the images.

The remainder of the paper is organized as follows. In Sec. \ref{sec2}, we provide a concise review of the Painlevé-Gullstrand form of the Lense-Thirring spacetime. In Sec. \ref{sec3}, we introduce the model of an accretion disk and the imaging method. Sec. \ref{sec4} presents our numerical simulation and results, accompanied by comprehensive discussions. Sec. \ref{sec5} provides a summary and discussion of our work. Additionally, we will work in units where $G=c=1$.

\section{Review of the Painlev\'e-Gullstrand form of the Lense-Thirring spacetime}\label{sec2}

The standard Lense-Thirring metric, a century-old approximation, describes the gravitational field surrounding a rotating mass at a large distance. It relies solely on the total mass and angular momentum $J$ of the source. While it doesn't precisely solve the vacuum Einstein equations, it does asymptotically approach the Kerr metric over large distances. Recently, Baines and colleagues developed an explicit Painlevé-Gullstrand version of the Lense-Thirring spacetime \cite{Baines:2020unr}, where the metric is defined as follows:

\bea\label{metric}
ds^2=-dt^2+\left(dr+\sqrt{\frac{2M}{r}}dt\right)^2+r^2\left[d\theta^2+\sin^2\theta\left(d\phi-\frac{2J}{r^3}dt\right)^2\right]\,.
\eea

For the slow rotation approximation solution, it's evident that as the angular momentum $J$ approaches zero, the solution simplifies to a Schwarzschild solution. The revised Lense-Thirring metric is in a unit-lapse form, which features a "rain" geodesic, making the physical interpretation of these spacetimes particularly straightforward and elegant. In \cite{Baines:2020unr}, it's demonstrated that the Painlevé-Gullstrand interpretation of Lense-Thirring spacetime is of Petrov type I, but it still retains many beneficial characteristics in its geodesic. As for the Carter constant $C$, the fourth constant in solving geodesics, \cite{Baines:2021qaw} introduces the Killing vector in the Painlevé-Gullstrand interpretation of Lense-Thirring spacetime. Furthermore, \cite{Baines:2022srs} provides the constant-$r$ geodesics in the Painlevé-Gullstrand form of Lense-Thirring spacetime, with results that are qualitatively similar to those for Kerr spacetime.

Although the Lense-Thirring metric is not an exact solution to Einstein's equations, this solution is approximately valid when $J/r_s^2 << 1$, where $r_s$ is the radius of the stellar surface \cite{mashhoon1984influence}. This effectively describes the external spacetime of a slowly rotating celestial body, which notably lacks an event horizon. Therefore, we propose using the Painlevé-Gullstrand version of the Lense-Thirring spacetime to represent the external spacetime of a CO. It is important to note that the metric in Eq. (\ref{metric}) exhibits a coordinate singularity at $r_h=2M$, implying that the stellar surface radius must be greater than $2M$. Therefore, in subsequent computations, similar to our previous work \cite{Wang:2023nwd}, we shall focus on the conditions $J/r_s^2 << 1$ and $r_s>2M$.

Next, we delve into the geodesics of the Painlevé-Gullstrand form of the Lense-Thirring spacetime, with further details available in \cite{Baines:2021qfm}. This spacetime reveals a non-trivial Killing tensor $K_{ab}$, and for any affine parameter, there is a conserved quantities Carter constant $C$:
\be
C=K_{ab}\left(\frac{\partial}{\partial\lambda}\right)^a\left(\frac{\partial}{d\lambda}\right)^b=r^4\left[\left(\frac{d\theta}{d\lambda}\right)^2+\sin^2\theta\left(\frac{d\phi}{d\lambda}-\frac{2J}{r^3}\frac{dt}{d\lambda}\right)^2\right]\,,
\ee
where $\lambda$ is the affine parameter along a geodesic. Therefore, for a geodesic of a free particle, where the four-momentum can be denoted as $p^a$, we can derive the equation of motion for the particle from the four conserved quantities associated with the free particle: (1)  the energy $E=-p_t$, (2) the angular momentum $L=-p_\phi$, (3) the mass $g_{ab}p^ap^b=m^2$, and (4) the Carter constant $C$. Without loss of generality, we set $m=0$ for the mass of the null geodesic and $m=-1$ for the timelike geodesic. Then, using these four quantities, the projections, the projections $\dot{t},\,\dot{r},\,\dot{\theta},\,\dot{\phi}$ of $p^a$ in the $t,\,r,\,\theta,\,\phi$ direction can be expressed as
\bea\label{geeq}
\dot{t}&=&\frac{E-2JL/r^3+S_r\sqrt{(2M/r)R(r)}}{(1-2M/r)}\,,\nn\\
\dot{r}&=&S_r\sqrt{R(r)}\,,\nn\\
\dot{\theta}&=&S_\theta\frac{\sqrt{\Theta(\theta)}}{r^2}\,,\nn\\
\dot{\phi}&=&\frac{L}{r^2\sin^2\theta}+2J\frac{E-2JL/r^3+S_\phi\sqrt{(2M/r)R(r)}}{r^3(1-2M/r)}\,,
\eea
where we define
\bea
R(r)&=&\left(E-\frac{2JL}{r^3}\right)^2-\left(m+\frac{C}{r^2}\right)\left(1-\frac{2M}{r}\right)\,,\\
\Theta(\theta)&=&C-\frac{L^2}{\sin^2\theta}\,,
\eea
as the effective potential functions governing the radial and polar motions. Here, $S_r=\pm1$ denotes outgoing and ingoing geodesics, $S_{\theta}=\pm1$ represens increasing and decreasing declination geodesic, and $S_{\phi
}$ signifies prograde and retrograde geodesics.

We know that the null geodesic, which satisfies $R(r)=\partial_r R(r)=0$, is the spherical photon orbit. Spherical photon orbits are closely related to the black hole shadow curves, and we discussed their relationship in the Painlevé-Gullstrand form of the Lense-Thirring space-time in our previous work \cite{Wang:2023nwd}. Next, we would like to introduce the timelike circular orbit, which satisfies $R(r)=\partial_r R(r)=0$ and $\theta=\pi/2$. By calculating the above formulas, we can obtain the conserved quantities $E_{\text{cir}}$ and $L_{\text{cir}}$ of the timelike circular orbit. The stable timelike circular orbit satisfies $\partial^2_r R(r)\leq0$. When $\partial^2_r R(r)=0$, the corresponding radius is the radius of the innermost stable circular orbit (ISCO), that is, $r_{\text{isco}}$. In the Painlevé-Gullstrand form of the Lense-Thirring spacetime, we have
\be
r_{\text{ISCO}_\pm}=6M\pm\frac{4\sqrt{2}}{\sqrt{3}}\frac{J}{M}+\mathcal{O}(J^2),
\ee
where the “$\pm$” symbols separately denote the prograde and the retrograde innermost timelike circular orbit, respectively. However, in subsequent calculations, we will not rely on this approximate expression. Instead, we solve directly to obtain more precise numerical outcomes.

\section{ Accretion disk and imaging method}\label{sec3}

In this section, we will concentrate on the accretion disk model and the methodology for capturing images of the CO illuminated by the accretion disk.

\begin{figure}[h!]
\centering
\includegraphics[width=0.9\textwidth]{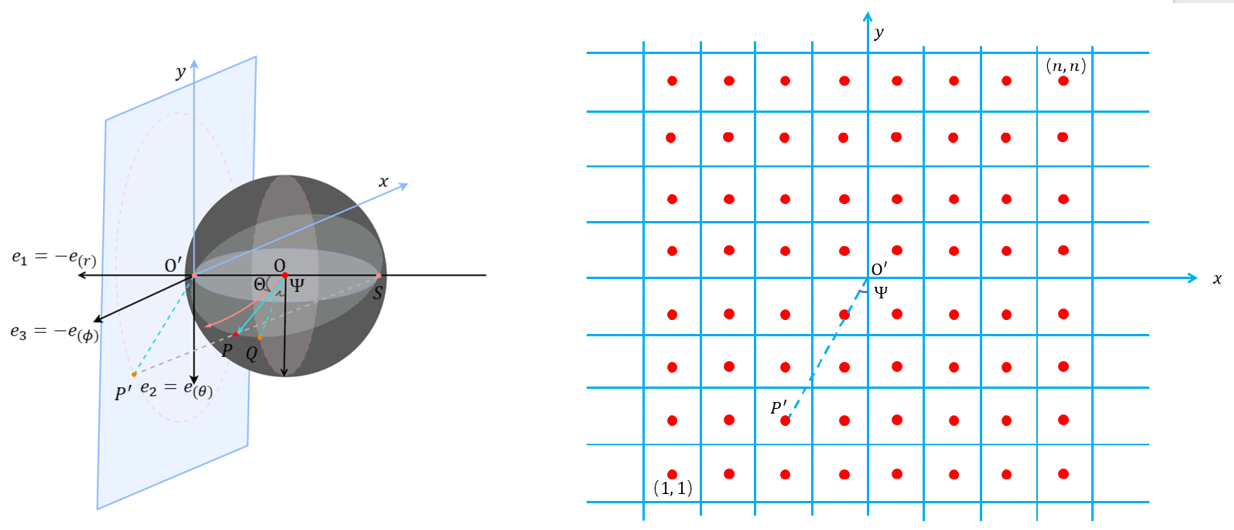}
\caption{Both of these figures are derived from our previous work \cite{Hu:2020usx}. In the left panel of the figure, we illustrate the stereographic projection method. The angle between the tangent of a geodesic line in three-dimensional space and the ZAMO frame is represented by the celestial coordinates $(\Theta,\,\Psi)$. In the right panel of the figure, we present the observation screen used in the ray-tracing method. The observation screen is divided into an $n\times n$ pixel grid.}
\label{sy}
\end{figure}

The Painlevé-Gullstrand form of the Lense-Thirring spacetime is stationary and axisymmetric. Thus, for an observer, their position can be denoted as $(0,\,r_o,\,\theta_o,0)$. We will proceed under the assumption that the observer is situated within the frame as follows:
 \begin{eqnarray}
e_0&=&\hat{e}_{(t)}=\partial_t-\sqrt{\frac{2M}{r}}\partial_r+\frac{2J}{r^3}\partial_\phi\,,\\
e_1&=&-\hat{e}_{(r)}=-\partial_r\,,\\
e_2&=&\hat{e}_{(\theta)}=\frac{1}{r}\partial_\theta\,,\\
e_3&=&-\hat{e}_{(\phi)}=-\frac{1}{r\sin\theta}\partial_\phi\,.
\label{tr}
 \end{eqnarray}
This corresponds to the "rain" geodesics in the Painlevé-Gullstrand form of the Lense-Thirring spacetime metric. It is straightforward to confirm that these bases are normalized and orthogonal to each other. Furthermore, given that $\hat{e}_{(t)}\cdot\partial_\phi=0$, the observer with the 4-velocity $\hat{u}=e_0$ possesses zero angular momentum. Consequently, this frame is typically referred to as the Zero angular momentum observer (ZAMO) reference frame. We can expand the four-momentum of photons in the ZAMO coordinate system as follows:
\be
p_{(\mu)}=k_\nu e^\nu_{(\mu)},
\ee 
where $e^\nu_{(\mu)}$ is given in Eq(\ref{tr}) and $k^\mu$ is the four-momentum of the photon. By introducing celestial coordinates, we can obtain the projections on the observer's screen. We establish a Cartesian coordinate system on a square observation screen. The $x$-axis of this coordinate system is parallel to $e_{(\phi)}$, the $y$-axis is parallel to $e_{\theta}$, and the origin aligns with the ZAMO frame . In this study, we utilize the stereographic projection method for imaging on the observation screen. Specific details can be found in our previous work\cite{Hu:2020usx}. In the stereographic projection method, the relationship between Cartesian coordinates $(x, \,y)$ and four-momentum can be derived using the celestial coordinates $\Theta$ and $\Psi$. This relationship is illustrated in the left panel of Fig. \ref{sy} from our previous work\cite{Hu:2020usx}. The celestial coordinates can be expressed in terms of the four-momentum as follows:
\be
\cos\Theta=\frac{p^{(1)}}{p^{(0)}},\,\,\,\,\,\,\tan\Psi=\frac{p^{(3)}}{p^{(2)}}.
\ee
Furthermore, the Cartesian coordinates $(x,\,y)$ on the observation screen can be expressed as:
\be
x=-2\tan\frac{\Theta}{2}\sin{\Psi},\,\,\,\,y=-2\tan\frac{\Theta}{2}\cos{\Psi}.
\ee

A point $(x,\,y)$ on the observation screen uniquely corresponds to the four-momentum $p_\mu$ of a photon. The observation screen is divided into an $n \times n$ pixel grid, as depicted in the right panel of Fig. \ref{sy}. In our work, we employ the ray-tracing method to numerically evolve the geodesics starting from the pixel points, directed towards the black hole. This allows us to generate the accretion disk image of CO.

Next, we will introduce the geometrically thin and optically thin accretion disk model we employed. The accretion disk consists of free electrically neutral plasmas that travel along the geodesics on the equatorial plane. The width of the accretion disk is significantly larger than that of the ISCO, and the internal boundary of the accretion disk extends to the surface of the CO. When the accretion flow is within the ISCO, it moves along critical plunging orbits, and there is a radial velocity component $u^r$ in addition to the $\phi$ component of velocity. However, outside the ISCO, the plasmas move along circular orbit geodesics. Hence, the four-momentum per unit mass can be expressed as:
\be
p^\mu=\zeta(1,\,0,\,0,\,\Omega(r)),\,\,\,\,\zeta=\sqrt{\frac{-1}{g_{tt}+2g_{t\phi}\Omega(r)+g_{\phi\phi}\Omega^2(r)}},\,\,\,\,\Omega=\frac{p^{\phi}(r)}{p^t(r)}=\frac{p^\phi}{\zeta}\,.
\label{four}
\ee

Next, we turn our attention to the intensity of the CO image illuminated by the accretion disk. Clearly, the intensity observed on the screen is associated with the interaction between the photons reaching the screen and the accretion disk. For simplicity, we ignore the reflection of the accretion disk. The change in intensity can be computed using the subsequent formula \cite{Davis:2019xee}:
\be
\frac{d}{d\lambda}\left(\frac{I_\nu}{\nu^3}\right)=\frac{J_\nu-\kappa_\nu I_\nu}{\nu^2}, 
\ee
In this equation, $\lambda$ denotes the affine parameter of null geodesics, while $I_\nu$, $J_\nu$, $\kappa_\nu$ represent the specific intensity, the emissivity, and the absorption coefficient at frequency $\nu$, respectively. In the scenario where light propagates in a vacuum, both the emissivity $J_\nu$ and the absorption coefficient $\kappa_\nu$ are $0$. This scenario infers that the ratio $I_\nu/\nu^3$ remains constant along the geodesics. 

As we assume that the disk is geometrically thin, the radiation and absorption coefficients can be considered constant during the process of light passing through the disk. In the ray-tracing method, the relationship between $I_n$ and $I_{n-1}$ can be derived by integrating the aforementioned equation. Ultimately, the intensity of a pixel on the observer screen can be calculated as:
\be
I_{\nu_o}=\sum^{N_{max}}_{n=1}\left(\frac{\nu_o}{\nu_n}\right)^3\frac{J_n}{\tau_{n-1}}\left[\frac{1-e^{-\kappa_n \nu_n\Delta\lambda_n}}{\kappa_n}\right],
\label{tau}
\ee
where the subscript ``$n$"is used to signify the $n$-th time the photon passes through the accretion disk. $\nu_0$ is the frequency measured in the ZAMO frame, while $\nu_n$ is the frequency measured in the local co-moving frame with respect to the plasmas. $\Delta\lambda_n$ is the change in the affine parameter of the photon when passing through the disk for the $n$-th time. In Eq. (\ref{tau}), there is a quantity $\tau_m$,which represents the optical depth of photons emitted at the $m$-th pass through the accretion disk. It is used to describe the light absorption by the disk:
\be
\tau_m=
\begin{cases}
\text{exp}\left[\sum^m_{n=1}\kappa_n \nu_n\Delta\lambda_n\right]\,,\,\, \,\,\,\text{if}\,\,m\ge1,\\
1\,,\,\,\,\,\,\,\,\,\,\,\,\,\,\,\,\,\,\,\,\,\,\,\,\,\,\,\,\,\,\,\,\,\,\,\,\,\,\,\,\,\,\,\,\,\,\quad\quad\text{if}\,\, m=0\,.
\end{cases}
\ee
Specific details are discussed in our previous work \cite{Hou:2022eev}. We assume the disk is optically thin, which means the absorption of the disk is approximately zero. As a result, Eq. (\ref{tf}) can be simplified as:
\be
I_{\nu_o}=\nu_o\sum^{N_{\text{max}}}_{n=1} \Delta\lambda_ng_n^2J_n\,,
\label{tf}
\ee
where $g_n$ is the redshift factor, $I_{\nu_o}$ is the intensity of photons on the observation screen. Similar to the first black hole image of M87 released by EHT, we take $\nu_o$ as $230$ MHz. For simplicity, we set the emissivity to be:
\be\label{jn}
J_n=\frac{1}{r_n^3}\,.
\ee
Here, $r_n$ is the value of coordinate $r$ when the light ray crosses the disk for the $n$-th time. Within the framework of the Painlevé-Gullstrand form in Lense-Thirring spacetime, we have the analytical capability to calculate $\Delta\lambda_n$,
\be\label{lam}
\Delta\lambda_n=\int^{\frac{\pi}{2}+\Delta\theta}_{\frac{\pi}{2}-\Delta\theta}d\lambda\Bigg|_{r=r_n}
\simeq 2r_n^2\int^{\frac{\pi}{2}}_{\frac{\pi}{2}-\Delta\theta}\frac{d\theta}{\sqrt{C-\frac{L^2}{\sin^2\theta}}}=\frac{2r_n^2}{\sqrt{C}}\arcsin\left(\frac{\sin\Delta\theta}{\sqrt{1-\frac{L^2}{C}}}\right)\simeq \frac{2r_n^2\Delta\theta}{\sqrt{C-L^2}}\,.
\ee
We should note that the `$\simeq$' in Eq. (\ref{lam}) are employed under the assumption that the disk is geometrically thin. For a photon crossing the disk at a radius $r$, the disk is presented with an open angle $\theta_{\text{disk}}=2\Delta\theta$ around the equatorial plane. The redshift factor, denoted by $g_n$, for photons that have traversed the disk $n$ times, can be formulated as:  
\be
g_n=\frac{e}{\zeta_n(1-\Omega(r_n)l)}, \,\,\,\,\,\text{if}\,r\geq r_{\text{ISCO}},
\label{rsh}
\ee
where we have introduced the following definitions 
\be
l=\frac{\mathcal{L}}{\mathcal{E}}=\frac{k_\phi}{-k_t},\,\,\,\,e=\frac{\mathcal{E}_o}{\mathcal{E}}=\frac{p_{(0)}}{k_t},\,\,\,\,\zeta_n=\sqrt{\frac{-1}{g_{tt}+2g_{t\phi}\Omega_n+g_{\phi\phi}\Omega_n^2}},
\ee
to simplify Eq. (\ref{rsh}). Here, $l$ is the impact parameter of photons, $\mathcal{E}=-k_t$ and $\mathcal{L}=k_\phi$ are the conserved quantity of photons along the geodesic, and $e$ is the ratio of the observed energy on the screen to the conserved energy along a null geodesic. Note that in the Painlev'e-Gullstrand form of the Lense-Thirring spacetime, which is asymptotically flat, $e=1$ when $r_o\rightarrow\infty$. On the other hand, within the ISCO, the expression for the redshift factor $g_n$ is
\be\label{gni}
g_n=\frac{1}{u_{rn}(g^{rr} k_r-g^{r\phi} l)+E_{\text{ISCO}}(g^{tt}-g^{t\phi}l)+L_{\text{ISCO}}(g^{\phi\phi}l-g^{t\phi}))},\,\text{if}\, r\leq r_{\text{ISCO}},
\ee
where $u_{rn}$ is the radial velocity of the accretion flow at $r_n$.

\section{Numerical  simulation and results }\label{sec4}

By combining Eqs. (\ref{tf}), (\ref{jn}), (\ref{lam}), (\ref{rsh}), and (\ref{gni}), we can use a numerical technique \cite{Hu:2020usx, Hou:2022eev}, specifically backward ray-tracing, to generate images of the accretion disk surrounding a CO. It's important to note the presence of a constant factor $\nu_o \Delta \theta$ in the equations. We consider a fixed observing frequency of 230 GHz, which results in significant numerical values. Given the geometric thinness of the accretion disk and our primary focus on the variation in the image's brightness, we can conveniently set $\nu_o \Delta \theta=1$ during the numerical computations. 

In addition,  for simplicity without losing generality, we set the mass of the CO to $M=1$. Considering a distant observer, we thus define the observational distance as $r_o=1000$ and the field of view angular as $\alpha_{\text{fov}}=\frac{\pi}{160}$.  In our model, the inner boundaries of the accretion disk extend to the surface of the central object, while we set the outer limit at $r_{\text{out}}=20$. Given that the radiation varies inversely with $r^3$, the influence of radiation from areas distant from the  inner boundaries of the accretion disk around the CO can be effectively ignored. Hence, the choice of the outer radius is well justified. Recall the essential condition that $J/r_s^2 << 1$. Consequently, following our previous work \cite{Wang:2023nwd}, we choose $J=0.5$, with $r_s \ge 2.24$. At this stage, it can be inferred that the radial coordinates of the photon region meet the condition $2.47\simeq r_{p-}\le r\le r_{p+}\simeq 3.56$, where $r_{p\pm}$ represent the outer and inner light rings on the equatorial plane. In this study, our primary aim is to investigate whether the images of accretion disks exhibit photon ring structures when the spacetime has a complete photon region, a partial photon region, or no photon region for CO. If such structures exist, we seek to determine their completeness. Considering that the presence and location of the photon ring rely solely on the intersections of light trajectories crossing the equatorial plane multiple times, irrespective of the specific dynamics of the flow at that position, we can infer that the existence and position of the photon ring remain invariant, regardless of the flow direction. As a result, our analysis will be exclusively focused on the results from the prograde accretion flow.

\subsection{Photon ring}

Subsequently, we will consider four distinct values for the surface radius, denoted as $r_s=2.24+0.8i$, $i=0,\,1,\,2,\,3$. Each value of radius  correspondingly belongs to  $r_h<r_s<r_{p-}$, $r_{p-}<r_s<r_{p+}$, $r_{p+}<r_s<r_{\text{ISCO}+}$ and $r>r_{\text{ISCO}+}$, where $r_{\text{ISCO}+}\simeq 4.28$ represents the radius of the prograde ISCO. For the observational angles, we will consider two scenarios: one where $\theta_o=17^\circ$, and the other where $\theta_o=80^\circ$.

\begin{figure}[h!]
\centering
\includegraphics[width=0.9\textwidth]{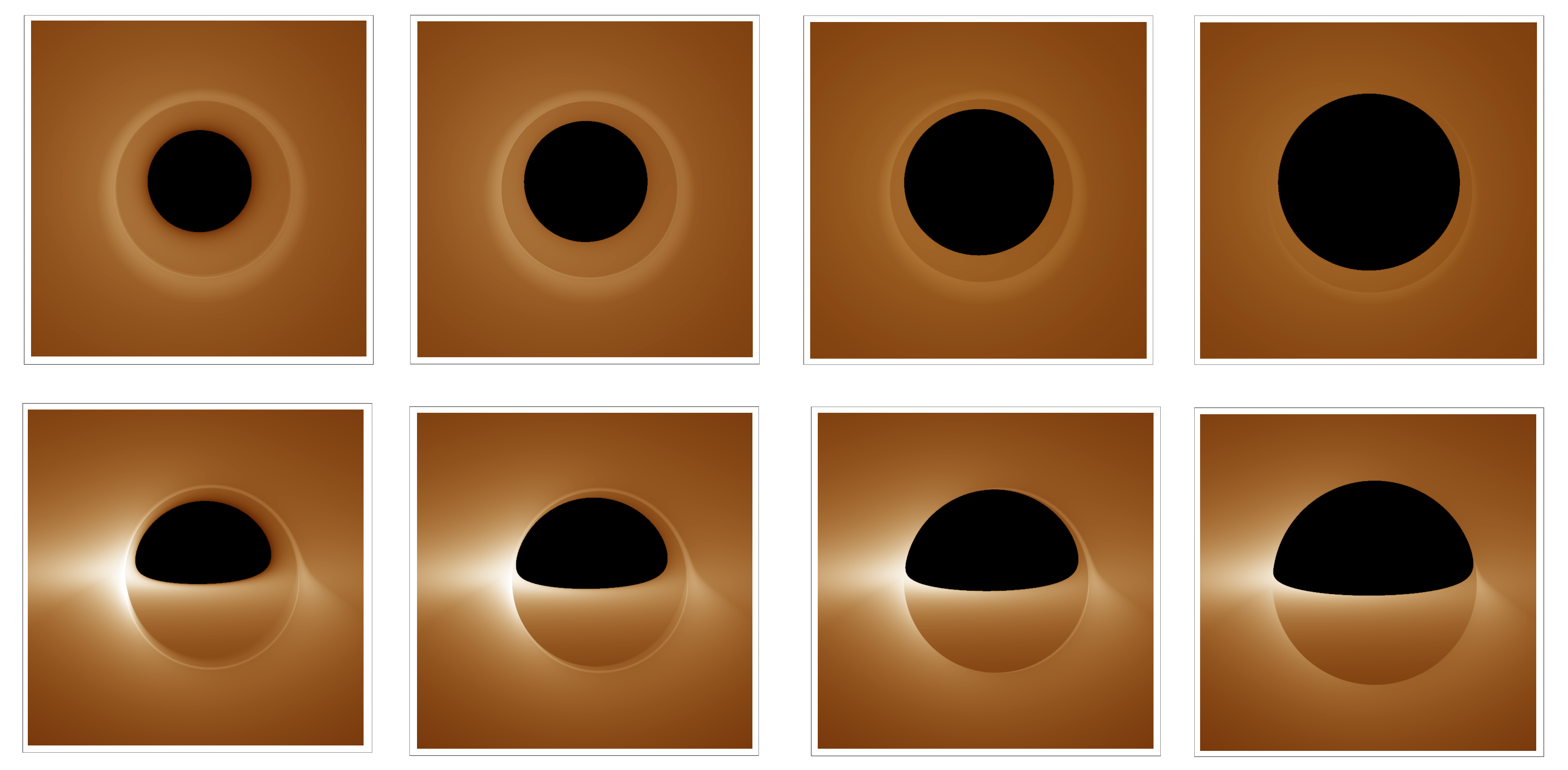}

\caption{Images of prograde accretion flows around a CO. The top row of the images is
observed at $\theta_o=17^\circ$, while the bottom row showcases observations made at $\theta_o=80^\circ$. From left to right, the images represent $r_s=2.24+0.8i$, where $i$ takes values of $0,\,1,\,2,\,3$.}
\label{proi}
\end{figure}

\begin{figure}[h!]
\centering
\includegraphics[width=0.9\textwidth]{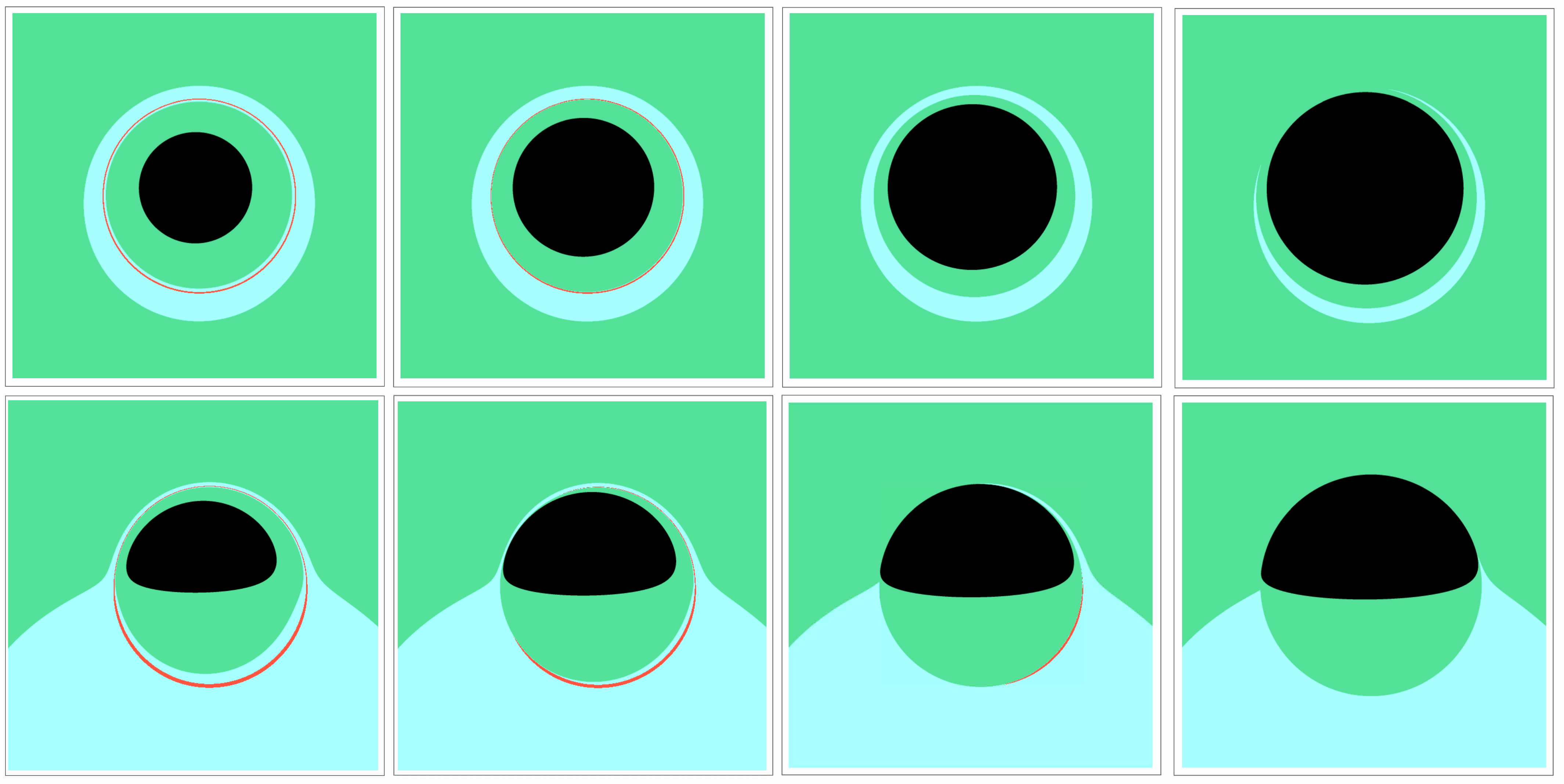}

\caption{Visualizations depict the range of frequencies at which light rays intersect the equatorial plane in black hole images. Pale green signifies a single intersection, representative of primary images; light blue indicates double intersections, corresponding to lensed images; and red denotes three or more intersections, linked to photon rings. The observational angles and the parameters of the surface radius maintain consistency with those detailed in Fig. \ref{proi}.}
\label{NN}
\end{figure}

In Fig. \ref{proi}, we present the images of a prograde accretion disk outside a CO, with the top row displaying results for $\theta_o=17^\circ$ and the bottom row for $\theta_o=80^\circ$. Progressing from left to right are the outcomes for $r_s=2.24+0.8i$, where $i$ takes on values of $0, 1, 2, 3$. Within the illustration, the manifestation of the inner shadow is conspicuously observable, aligning with expectations, as the celestial body can be conceived as non-luminous relative to the highly luminous accretion disk. Furthermore, the primary and secondary images of the accretion disk can be distinctly delineated in the diagram. However, given the typically narrow profile of the photon ring, if it indeed exists, it tends to be easily overwhelmed within the secondary images, impeding direct visibility. Therefore, in Fig. \ref{NN}, we have differentiated regions of different image levels based on the photon trajectories traversing the accretion disk, color-coded for distinction. Light green denotes primary images, light blue signifies secondary images, and red represents the photon ring. This depiction, unlike Fig. \ref{proi}, does not incorporate image intensities, facilitating a clearer resolution of the photon ring structure compared to Fig. \ref{proi}.

\begin{figure}[h!]
\centering
\includegraphics[width=0.9\textwidth]{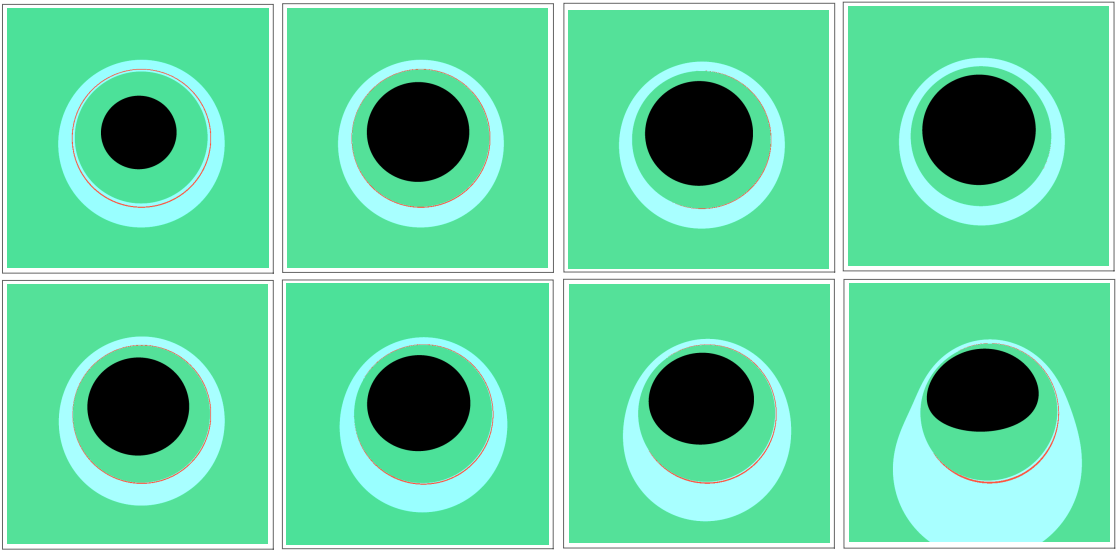}

\caption{Different selective outcomes are determined by the observation angle and surface radius. The first row showcases $\theta_o=17^\circ$, revealing outcomes for surface radii $r_s=2.01,\,3.08,\, 3.3,\, \text{and}\, 3.52$ progressing from left to right. In the subsequent row, with $r_s=3.04$ held constant, outcomes for $\theta_o=21^\circ,\,31^\circ,\,41^\circ,\, \text{and}\, 61^\circ$ are displayed from left to right.}
\label{nnc}
\end{figure}

We plan to defer the discussion related to the inner shadow to a later subsection. Our initial emphasis is primarily on the presence and integrity of the photon ring under various scenarios. In Figure \ref{NN}, it is observed that for $\theta_o=17^\circ$, a complete photon ring is present only when $r_s=2.24$ and $r_s=3.04$. However, for $r_s=3.84$ and $r_s=4.64$, the photon ring is completely absent. Focusing on $\theta_o=80^\circ$, we observe that a complete photon ring only exists at $r_s=2.24$. At $r_s=3.04$ and $r_s=3.84$, incomplete photon rings are present, with the ring at $r_s=3.84$ being noticeably more incomplete than the one at $r_s=3.04$. Furthermore, at $r_s=4.64$, the photon ring completely disappears. These observations suggest that as the surface radius $r_s$ increases for different observational angles, the complete photon ring undergoes a transition from weakening to partial disappearance, progressing to complete vanishing.

To further investigate this phenomenon, we have kept $\theta_o=17^\circ$ while systematically varying the surface radius $r_s$ to scrutinize the alterations in the photon ring. Conversely, with $r_s=3.04$ held steady, we have continuously adjusted the observational angle $\theta_o$ to monitor the changes in the photon ring. The crucial findings are illustrated in Fig. \ref{nnc}. In the first row, where $\theta_o=17^\circ$, the results for surface radii $r_s=2.01,\,3.08,\, 3.3\, \text{and}\, 3.52$ are presented from left to right. In the second row,while maintaining $r_s=3.04$, the results for $\theta_o=21^\circ,\,31^\circ,\,41^\circ\, \text{and}\, 61^\circ$ are exhibited from left to right. These results are highly representative. With the observational angle fixed at $\theta_o=17^\circ$, a critical surface radius of approximately $r_{sc}=3.08$ was observed. For $r_s<r_{sc}$, the photon ring remains intact. However, when $r_s>r_{sc}$, a segment of the photon ring begins to gradually disappear, until it completely vanishes for $r_s>3.52$. With the surface radius held constant at $r_s=3.04$, we identify a critical observational angle of approximately $\theta_{oc}=21^\circ$. When $\theta_o<\theta_{oc}$, the photon ring appears complete; as $\theta_o>\theta_{oc}$, the completeness of the photon ring decreases. Importantly, when $\theta_o>40^\circ$, the photon ring's shape attains a state of consistent imperfection. That is, although the photon ring is fragmented, the level of imperfection remains steady.

\begin{figure}[h!]
\centering
\includegraphics[width=0.6\textwidth]{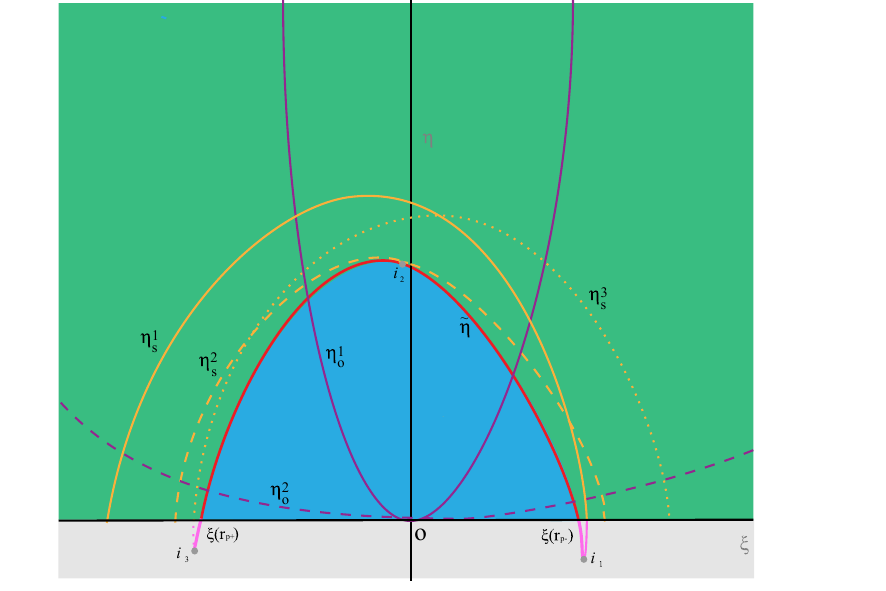}

\caption{The functions $\tilde{\eta}(\tilde{\xi})$, $\eta_o(\xi_o)$, and $\eta_s(\xi_s)$ are illustrated in the $\xi O \eta$ plane. Here, $\eta_o^1$ and $\eta_o^2$ represent the values of $\eta_o(\xi_o)$ at $\theta_o=17^\circ$ and $\theta_o=80^\circ$ respectively. Meanwhile, $\eta_s^1$, $\eta_s^2$, and $\eta_s^3$ denote the values of $r_s$ under the conditions $r_s<r_{p-}$, $r_{p-} < r_s < r_{p+}$, and $r_s > r_{p+}$ respectively. In particular, the curve $\eta_s^2$ intersects $\tilde{\eta}$ at a point $i_2$ near the $\eta$-axis, which we denote with a gray dot. Additionally, the pink line below the $\xi$-axis represents an extension of the domains for the functions $\tilde{\eta}$, $\eta_s^1$, and $\eta_s^3$, with imaginary extensions denoted as $i_1$ and $i_3$ at their respective hypothetical intersection points.}
\label{etaxi}
\end{figure}

This captivating pattern merits careful examination. For a more profound understanding of this phenomenon, we will explore the impact parameters of photons. Adhering to the convention established in our prior work \cite{Wang:2023nwd}, we define the impact parameters as follows:
\bea
\xi=\frac{L}{E}\,\quad\quad \eta=\frac{C-L^2}{E^2}\,.
\eea
As clarified in our previous work \cite{Wang:2023nwd}, pertaining to the shadow of the CO, there are three characteristic functions of $\eta$ with respect to $\xi$, namely, $\tilde{\eta}(\tilde{\xi})$, $\eta_o(\xi_o)$, and $\eta_s(\xi_s)$. $\tilde{\eta}(\tilde{\xi})$ describes the photon region, encompassing all spherical photon orbits, which can be determined by applying the radial potential function for null geodesics, resulting in the equation $R(r)=\partial_r R(r)=0$. $\eta_o(\xi_o)$ signifies photons that reach the observer precisely, achievable by setting $\Theta(\theta_o)=0$. Meanwhile, $\eta_s(\xi_s)$ represents photons whose turning points are exactly at the surface radius $r_s$, derivable via $R(r_s)=0$. The interplay among these three functions of $\eta$ in relation to $\xi$ can be intuitively understood through their graphical illustrations, as shown in Fig. \ref{etaxi}. In the graph, the red solid line represents $\tilde{\eta}$, with its intersection points on the $\xi$-axis corresponding to the radii $r_{p-}$ (left) and $r_{p+}$ (right). The function $\eta_s$ are depicted by yellow lines, where the solid line represents $\eta_s^1$, the dashed line represents $\eta_s^2$, and the dotted line represents $\eta_s^3$, corresponding to the conditions $r_s<r_{p-}$, $r_{p-} < r_s < r_{p+}$, and $r_s > r_{p+}$ respectively. The purple lines represent $\eta_o$, with the solid line being $\eta_o^1$ and the dashed line being $\eta_o^2$, representing the values at $\theta_o=17^\circ$ and $\theta_o=80^\circ$ respectively.

In preparation for subsequent discussions, it is vital to emphasize certain overarching properties of the characteristic functions depicted in Fig. \ref{etaxi}. Herein, the function plot of $\tilde{\eta}$ consistently lies below $\eta_s$. Particularly, when $r_{p-} < r_s < r_{p+}$, an intersection point exists above the $\xi$ axis between $\eta_s^2$ and $\tilde{\eta}$, as illustrated by $i_2$ in Fig. \ref{etaxi}. Extending our analysis to the other two distinct intervals, it becomes clear that $\eta_s^1$ and $\eta_s^3$ do not intersect with $\tilde{\eta}$ in the region where $\eta>0$. Delving into the case of $r_s < r_{p-}$, where $\eta_s^1 = 0$ corresponds to a larger $\xi$ value exceeding $\xi_{r_{p-}}$, it follows that a smaller $\xi$ value must necessarily be less than $\xi_{r_{p+}}$. Conversely, for $r_s > r_{p+}$, it is apparent that the smaller $\xi$ value corresponding to $\eta_s^3 = 0$ must be less than $\xi_{r_{p+}}$, while the larger $\xi$ value should exceed $\xi_{r_{p-}}$. Maintaining the fixed nature of $r_s$ for both $\eta_s^1$ and $\eta_s^3$, and leveraging the continuity of functions, we extend $\eta_s^1$, $\eta_s^3$, and $\tilde{\eta}$ to the realm of $\eta < 0$ as delineated by the pink lines in Fig. \ref{etaxi}. Consequently, it ensues that $\eta_s^1$ and $\eta_s^3$ intersect $\tilde{\eta}$ and $i_3$ in Fig. \ref{etaxi}, respectively. Notably, these pink lines and intersection points $i_1$ and $i_3$ are entirely hypothetical constructs designed for discussion purposes and do not align with physical reality.

In the aforementioned research \cite{Wang:2023nwd}, we conducted a thorough analysis of how the presence and attributes of these characteristic functions shape the shadow curve of the CO. Without reiterating the introduction, our current study is chiefly devoted to exploring the factors that lead to the disintegration and vanishing of photon rings. It is crucial to underscore that the definition of a photon ring relates to the image created when light rays intersect the equatorial plane three or more times before they reach the observer. Conversely, the critical curve denotes the image formed at the observer's position by photons that traverse spherical photon orbits and are capable of reaching the observer. Fundamentally, these light rays intersect with the equatorial plane numerous times prior to their arrival at the observer. Hence, the critical curve is an integral part of the photon ring, enveloped within the confines of the photon ring. 

\begin{figure}[h!]
\centering
\includegraphics[width=0.9\textwidth]{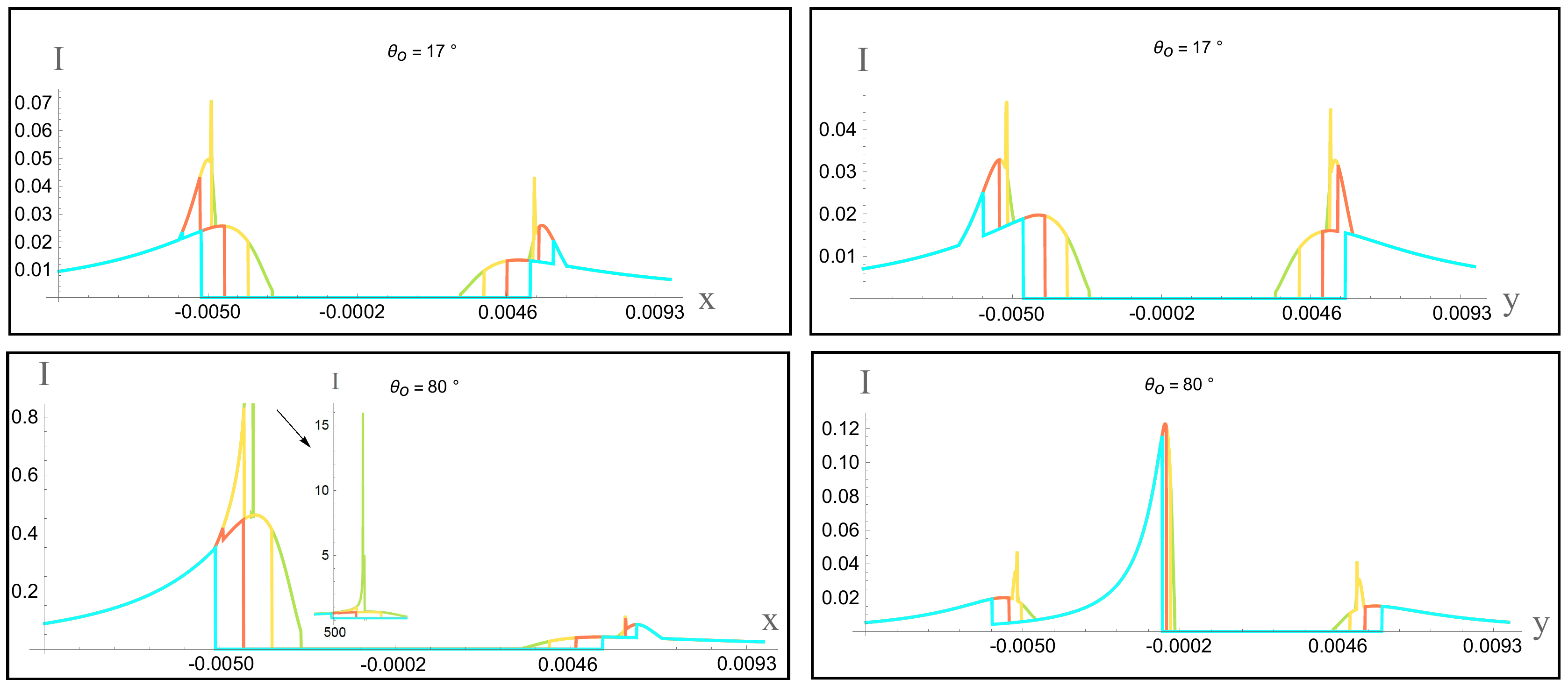}

\caption{The analysis encompasses the intensity cutss along the $x$-axis and $y$-axis of the accretion disk images, characterized by a prograde flow. The colors green, yellow, red, and blue correspond to $r_s=2.24$, $r_s=3.04$, $r_s=3.84$, and $r_s=4.64$, respectively.}
\label{pxy}
\end{figure}

Now, having established the necessary foundation, we are ready to elaborate on the phenomena illustrated in Figs. \ref{NN} and \ref{nnc} using Fig. \ref{etaxi}. Consider the situation where $r_s < r_{p-}$. In this case, the photon region exterior to the central object is fully formed, resulting in a closed critical curve irrespective of changes in the viewing angle. Therefore, under these circumstances, the photon ring is also guaranteed to be closed, demonstrating complete coherence without any interruptions. As we turn our attention to $r_{p-} < r < r_{p+}$, the situation becomes more complex. At this point, the CO partially obscures the photon region, resulting in a segment of $\tilde{\eta}$ to the right of $i_2$ corresponding to light that cannot reach the observer. As a result, the critical curve observed by the viewer is inevitably disjointed, displaying a partial break. However, this does not imply that the photon ring must also be disjointed. This is because the light corresponding to the portion to the right of $i_2$ in $\eta_s^2$ can indeed reach the observer. If this segment of $\eta_s^2$ aligns closely with $\tilde{\eta}$, indicating that the impact parameters of these photons closely match the critical impact parameters, then despite crossing the equatorial plane numerous times before reaching the observer, it remains conceivable for this light to intersect the equatorial plane three times or more, potentially forming a closed photon ring. From Fig. \ref{etaxi}, we can observe that when $\theta_o = 17^\circ$, the region where $\eta_s^2$ and $\tilde{\eta}$ reside is very close, while for $\theta_o = 80^\circ$, $\eta_s^2$ and $\tilde{\eta}$ are significantly distant from each other. Therefore, examining Fig. \ref{NN}, we find that at $r_s = 3.08$, the photon ring is complete for $\theta_o = 17^\circ$, but it is fragmented for $\theta_o = 80^\circ$. For $r > r_{p+}$, the photon region is completely obscured by the CO, making the critical curve invisible. However, the potential for a photon ring still exists. In the case of $\theta_o = 17^\circ$, where $\eta_s^2$ is significantly distant from $\tilde{\eta}$, we note from Fig. \ref{NN} that the last two images in the top row do not have a photon ring. On the other hand, for $\theta_o = 80^\circ$, when $r_s$ is not too large, as shown in Fig. \ref{etaxi}, near $\xi(r_{p+})$, the closeness is remarkable. Therefore, the third image in the second row of Fig. 4 displays a partial photon ring, while in the fourth image, due to $r_s$ being excessively large, the photon ring entirely vanishes.

Similarly, the results of the four images in the top row of Fig. \ref{nnc}, which illustrate $\theta_o = 17^\circ$ while varying $r_s$, are now easily understood, eliminating the need for further explanation. Regarding the results shown in the second row of Fig. \ref{nnc}, which involve the fixed parameter $r_s = 3.08$ and the modification of the observation angle $\theta_o$, we continue our examination by referring back to Fig. $\ref{etaxi}$. In this context, we focus on $\eta_s^2$, then follow the trend of $\eta_o$ in response to changes in $\theta_o$, thereby noting the distance between $\eta_s^2$ and $\tilde{\eta}$. By performing this analysis, we have successfully clarified the intriguing phenomena related to the variations of the photon ring as explained in Fig. \ref{NN} and Fig. \ref{nnc}.

\begin{figure}[h!]
\centering
\includegraphics[width=0.5\textwidth]{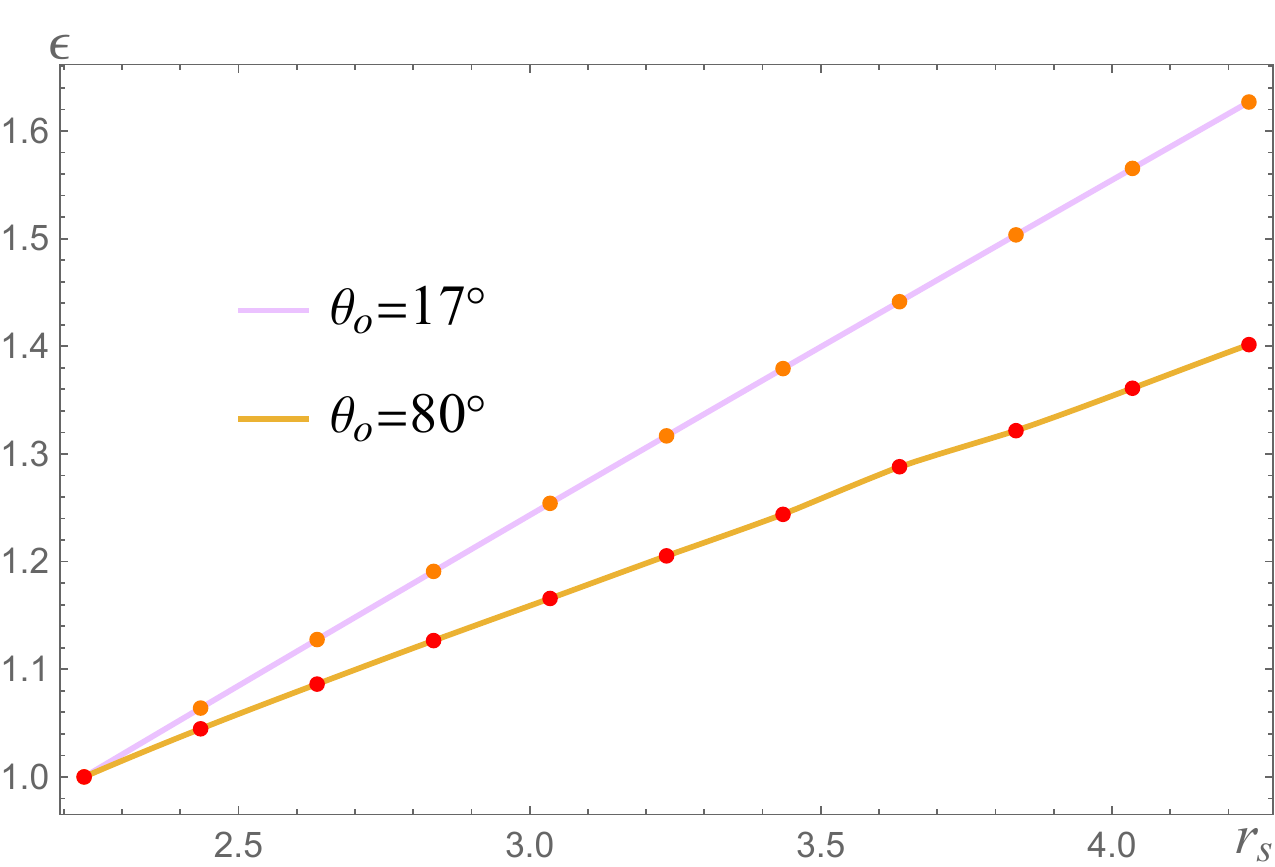}

\caption{Graphical representations of the function $\epsilon(r_s)$.}
\label{isc}
\end{figure}

\subsection{Intensity cut and inner shadow}

Although our study's main objective is to determine if the photon ring of a CO consistently appears as a closed structure, we will also examine the intensity cuts and inner shadows shown in the images to ensure a thorough analysis. 

 In Figure \ref{pxy}, the distribution of intensity along the $x$ and $y$ axes is displayed. We have marked four different surface radii, $r_s=2.24+0.8i$, where $i=0,\,1,\,2,\,3$, using curves of varying colors. The green curve represents $i=0$, yellow corresponds to $i=1$, red indicates $i=2$, and blue denotes $i=3$. In each image, the intensity in the central region is zero, and this region expands, aligning with the growth of the inner shadow as $r_s$ increases. For the result where $\theta_o=17^\circ$, the peaks of $r_s=2.24$ and $r_s=3.04$ correspond to the intensity of the photon ring. However, when $r_s=3.84$ and $r_s=4.64$, the peaks significantly decrease, mirroring the photon ring's disappearance in Fig. \ref{NN}. At $\theta_o = 80^\circ$, similar results are observed, except for a prominent peak in the $y$-axis profiles. This persistent peak, unlike the photon ring, is distinctive and separates the photon ring on both sides. This phenomenon is primarily due to the pronounced Doppler effect induced by the prograde accretion flow.

Finally, we shift our attention to discuss the variation of shadows with the surface radius $r_s$ of the CO, which refers to the inner shadow. Recalling our previous introduction to ZAMOs, on the observation screen, we can define the coordinates $(x_c,\,y_c)$ as
\be
x_c=\frac{\xmi+\xma}{2},\,\,\,\,y_c=\frac{\ymi+\yma}{2}\,,
\ee
These represent the geometric center of the inner shadow, where $x_{\text{max,\,min}}$ and $y_{\text{max,\,min}}$ are the maximal and minimal horizontal and vertical coordinates of the shadow boundary, respectively. We can transform the Cartesian coordinates into a polar coordinate system $(R,\,\psi)$ with $(x_c,\,0)$ as the origin, where $R=\sqrt{(x-x_c)^2+y^2}$. We define the average radius $\bar{R}$ as
\be 
\bar{R}=\int^{2\pi}_{0}\frac{R(\psi)}{2\pi}d\psi\,.
\ee
We introduce a parameter $\epsilon$ to characterize the changes in the inner shadow
\be
\epsilon=\frac{\bar{R}}{\bar{R}_0}-1\,,
\ee
here, $\bar{R}_0$ denotes the average radius of the inner shadow when $r_s=2.24$. In Fig. \ref{isc}, we take the surface radius of the CO $r_s=2.24+0.2(j-1),\,j=1,\,2,\,3...12$ and plot two curves of $\epsilon$ verse  $r_s$ at observation angles $\theta_o=17^\circ,\,\theta_o=80^\circ$. The pink curve corresponds to $\theta_o=17^\circ$, and the yellow curve corresponds to $\theta_o=80^\circ$. As illustrated in Fig. \ref{isc}, the inner shadows grow linearly with the increase of surface radius $r_s$ and the growth rate of $\epsilon$ at $\theta_o=17^\circ$ is significantly higher than that at $\theta_o=80^\circ$.

In Fig. 5 of our previous work \cite{Wang:2023nwd}, we illustrate the variation of the shadow curve for observation angles $\theta_o = 17^\circ$ and $\theta_o = 80^\circ$ as a function of the surface radius $r_s$. Upon comparison, we discern significant differences in the inner shadow outcomes. Specifically, the average radius of the shadow curve at $\theta_o = 80^\circ$ consistently exceeds that at $\theta_o = 17^\circ$ for a given $r_s$. Furthermore, for a fixed $r_s$, the differences between $\theta_o=17^\circ$ and $\theta_o=80^\circ$ in the inner shadow are much larger than those in the shadow curve. This suggests that it is more feasible to distinguish different observation angles when considering the inner shadow in observations. Additionally, the contrasting results for different observation angles on the shadow curve and inner shadow suggest that there is not a universal positive correlation between them.

\section{Summary}\label{sec5}

Continuing from our previous study \cite{Wang:2023nwd}, in this work, we utilized the Painlevé-Gullstrand form of the Lense-Thirring metric to model the spacetime outside the surface of the CO.We employed a geometrically and optically thin accretion disk as a light source and examined the images of the CO. We selected two different observational angles, $\theta_o=17^\circ,\,80^\circ$, and presented the accretion disk images of the CO with various surface radii, $r_s=2.04,\,3.04,\,3.84,\,\text{and}\, 4.64$. These values correspond to different regions: $r_h<r_s<r_{p-}$, $r_{p-}<r_s<r_{p+}$, $r_{p+}<r_s<r_{\text{pisco}}$ and $r>r_{\text{pisco}}$ respectively.

When a rotating celestial body is sufficiently compact, a photon shell forms around it. When the surface radius of the CO falls within the photon shell, as anticipated, the imaging results of the CO resemble those of a black hole, with observable closed photon ring structures from various angles. However, when the CO partially obscures the photon shell, we note an intriguing behavior. At an observation angle of $\theta_o = 17^{\circ}$, the resulting photon ring may display either a closed or broken structure, contingent on the size of the CO's surface radius. In most instances, at $\theta_o = 80^{\circ}$, only a broken photon ring structure is visible. When the CO entirely obscures the photon shell, the photon ring disappears completely at $\theta_o = 17^{\circ}$. However, at $\theta_o = 80^{\circ}$, it is still possible to discern a broken photon ring structure. These outcomes appear peculiar and seemingly lack a clear pattern. Yet, through the introduction and analysis of three characteristic functions related to photon impact parameters, namely $\tilde{\eta}$, $\eta_o$, and $\eta_s$, we offer a comprehensive explanation for the various structural states of the photon ring. Our findings not only predict the potential presence of broken photon ring structures or the possible absence of a photon ring theoretically, but also provide insights for observations of broken photon ring structures and non-existent photon ring structures as the EHT resolution improves.

Furthermore, we conducted an investigation into the inner shadow of the CO. Interestingly, we noticed substantial differences when compared to the CO's shadow curve. In particular, for an identical surface radius, the mean radius of the inner shadow at an observational angle of $\theta_o = 17^{\circ}$ exceeds the results at $\theta_o = 80^{\circ}$. This outcome contradicts the findings from the shadow curve \cite{Wang:2023nwd}. Additionally, the disparity in the mean radius of the inner shadow between $\theta_o = 17^{\circ}$ and $\theta_o = 80^{\circ}$ is markedly larger than the difference observed in the mean radius of the shadow curve. This suggests that if the center of the image captured by the Event Horizon Telescope EHT is a CO, the data derived from the inner shadow would be more appropriate for constraining the diverse physical parameters of the central celestial object.

\section*{Acknowledgments}
The work is partly supported by NSFC Grant Nos. 12205013, 12275004 and 12175008. MG is also endorsed by "the Fundamental Research Funds for the Central Universities" with Grant No. 2021NTST13. 

%\newpage
\bibliographystyle{utphys}
\bibliography{slow2}

\end{document}